\documentclass[twocolumn,showpacs,prb]{revtex4}
\usepackage{graphicx}
\usepackage{epsfig}
\usepackage{amsmath}
\usepackage{amssymb}
\usepackage{gensymb}

\newcommand{\beq}{\begin{eqnarray}}
\newcommand{\eeq}{\end{eqnarray}}

\begin{document}

\title{Buckling induced Zener polaron instability in half-doped manganites}

\author{Paolo Barone$^{1}$, Silvia Picozzi$^{1}$ and Jeroen van den Brink$^{2}$}

\affiliation{
$^1$CNR-SPIN 67100 L'Aquila, Italy\\
$^2$Institute for Theoretical Solid State Physics, IFW Dresden, 01171 Dresden, Germany 
}

\date{\today}

\begin{abstract}
By calculating the electronic, orbital, magnetic and lattice structure of half-doped manganites we establish the central
role of oxygen buckling caused by the tilting of the MnO$_6$ octahedra, which is, in essence, a steric effect. The buckling itself
does not change the Mn-Mn bonds. Instead, it drives the system toward an instability where these bonds can dimerize. In the
presence of electron-electron interactions, this instability can fully develop and beyond a critical buckling a Zener
polaron groundstate with dimerized spins, lattice, and orbitals forms spontaneously, resulting in a multiferroic state.
\end{abstract}

\pacs{75.47.Lx,71.10.Fd, 75.85.+t, 75.25.Dk}
%75.47.Lx Magnetic oxides
%71.10.Fd Lattice fermion models
%75.85.+t Magnetoelectric effect, multiferroics
%75.25.Dk Orbital, charge, and other orders, including coupling of these orders 
%75.50.Ee Antiferromagnetics
%77.80.-e Ferroelectricity and antiferroelectricity
%71.70.Ej       Spin-orbit coupling, Zeeman and Stark splitting, Jahn-Teller effect 

\maketitle

Materials where several degrees of freedom, spin, charge, orbital and lattice, are simultaneously active display very rich phenomenologies and exotic ground states, including ferromagnetic metals or antiferromagnetic insulators with complex charge and orbital ordered states~\cite{dagotto.rev}. Manganites clearly belong to this class, as is evident from the observed colossal response to external fields of magnetic, electric or strain origin~\cite{colossalm,colossale,colossals}, which indeed reveals a subtle balance between different interactions. Despite the enormous experimental activity and the considerable theoretical progress in understanding their rich phenomenology, manganites continue to unveil unexpected and fascinating properties like the coexistence of spontaneous ferroelectric polarization and magnetic order~\cite{kimura2,dagotto2,picozzi}, which defines them as potential multiferroics, offering new opportunities for, e.g., multifunctional magnetoelectric devices~\cite{eerenstein,khomskii}.

In particular doped rare-earth manganites, $R_{1-x}A_x$MnO$_3$ ($R$ and $A$ are rare-earth and alkaline-earth cations), were
recently suggested to have a potentially large multiferroic coupling~\cite{efremov}, particularly the half-doped
Pr$_{1/2}$Ca$_{1/2}$MnO$_3$ (PCMO)~\cite{jooss,colizzi} and  La$_{1/2}$Ca$_{1/2}$MnO$_3$ (LCMO)~\cite{gianluca}. Apart from
the possible appearance of a multiferroic state, the physical properties in the low-temperature regime of these compounds
are still debated. It is widely accepted that the local Mn magnetic moments can order  in the so-called CE-type
antiferromagnetic (AFM-CE) configuration characterized by double-zigzag spin chains~\cite{wollan} ($T_{\rm c}=175/155$ K for
PCMO/LCMO), but the picture is less clear for the charge (CO) and orbital ordering (OO) that are observed and two possible
scenarios have been envisaged. On the one hand a checkerboard pattern (CB) of alternating Mn$^{4+}$ and orbital-ordered
Mn$^{3+}$ has been proposed in a low-symmetry $P2_1/m$ space group~\cite{radaelli}. On the other hand,
a Zener polaron picture (ZP) has been suggested in the low-symmetry $P2_1nm$ space group, where equivalent Mn$^{3+}$ ions couple into ferromagnetic dimers, displaying an OO pattern with orbital states aligned along the same direction within each dimer. In this picture the CO originates from holes on oxygen atoms that bridge the Mn dimers~\cite{jooss,daoud}. The symmetry of the ZP state with $P2_1nm$ structure allows for a spontaneous electric polarization~\cite{jooss}, which is related microscopically to the presence of inequivalent oxygen sites~\cite{colizzi,gianluca}.

From the theoretical point of view, the CB picture has received considerable attention in recent years and has been
investigated by means of computational and mean-field analysis of realistic models~\cite{dagotto.rev}. The AFM-CE
configuration was shown to give rise to a band-insulating behavior that emerges from the topology of the zigzag
chains in the presence of large Hund coupling between $t_{2g}$ Mn core spins and $e_g$
electrons~\cite{dagotto.rev,brink,hotta}, which is also responsible for the emerging orbital-ordering pattern. The
checkerboard CO pattern can then come about via either electron-electron~\cite{brink} or electron-lattice
Jahn-Teller~\cite{hotta} interactions. On the other hand, the ZP state has been studied mainly in the framework of
first-principles calculations~\cite{colizzi,gianluca,ferrari,patterson} due to its complex nature and because of the expected importance of the oxygens ions, which are usually not included in model Hamiltonian approaches to manganites.

In this Brief Report we consider a straightforward extension of the degenerate double-exchange (DDEX) model that includes
the oxygen-buckling mode modulating the
Mn-Mn hopping processes by changing the Mn-O-Mn angle $\varphi$. Such a buckling
is ubiquitous in perovskites and is induced by a tilting and rotation of the
MnO$_6$ octahedra. The amount of tilting depends on the ionic radius of the rare-earth cation that is used and is, in
essence, a steric, and thus very robust,
effect. We find that the interplay of this buckling mode with the
electron-electron ($e$-$e$) interaction generates an instability toward Mn-Mn
dimerization. The emerging dimerized state has the qualitative features predicted for the ZP state, in particular, the same
OO pattern and a strongly reduced charge disproportion between Mn ions. The inclusion of further electron-lattice ($e$-$l$)
interactions, in particular, the coupling with the Jahn-Teller (JT) mode, further stabilizes this phase.  We analyze the different contributions to the ferroelectric polarization of the multiferroic state that emerges.

\begin{figure}
\includegraphics[width=8.6cm]{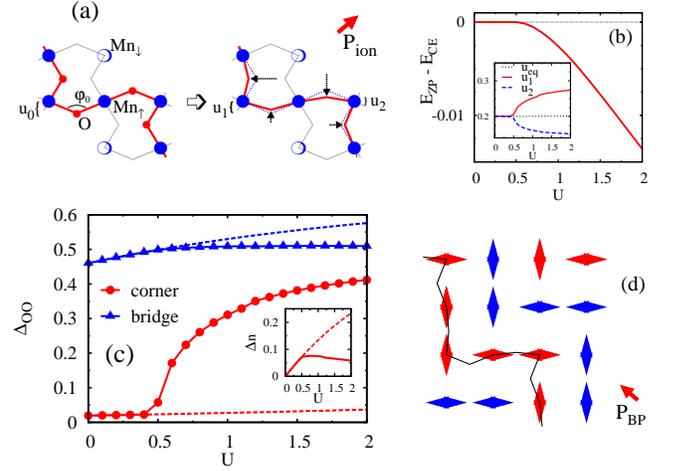}
\caption{(a) Schematic representation of the buckling distortion in the one-dimensional zigzag chain and of the dimerization effect induced by the local interaction (manganeses with opposite spins belong to different chains). (b) Energy gain of the dimerized (ZP) state with respect to the non-dimerized (CE) state as a function of $U$ (at $E_{JT}=0$) for an initial distortion $u_0=0.3$. Inset: equilibrium distortion $u_{eq}$ for the non-dimerized state compared to the staggered one found in the ZP state. (c) Difference of the on-site orbital occupancy between the most and least occupied orbitals as a function of $U$ in the dimerized state, on corner (circle) and bridge (triangle) sites. Inset: charge disproportion between corner and bridge Mn sites. Dotted lines represent the same quantities in the non-dimerized state. (d) Schematic representation of the orbital-order pattern in the dimerized state. }
\label{fig1}
\end{figure}

{\it Model.} In the infinite Hund coupling approximation,  $e_g$ electrons effectively only hop between sites with
ferromagnetically aligned core spins. As a consequence in the AFM-CE phase hopping is allowed merely within one-dimensional
zigzag chains of parallel core spins, changing direction as $\{..x,x,y,y..\}$. In the model Hamiltonian we also include the effect of the GdFeO$_3$-type distortions, a change of the Mn-O-Mn angle $\varphi$, on the hopping parameters as well as the $e$-$e$ and $e$-$l$ interactions:
\beq\label{ham}
H &&= -\sum_{\langle ij\rangle\alpha\beta}\, t_{\alpha\beta}^{ij}\,d^{\dagger}_{\alpha i}d^{\phantom{\dagger}}_{\beta j}\,+\,
U\,\sum_i \, n_{ia}\,n_{ib} \,+ \\
&& \lambda \sum_i \left(\,Q_{1i}\,n_i\, +\,Q_{2i}\tau^x_{i}\,+\,Q_{3i}\tau^z_{i}\right)
\,+\,\frac{1}{2}\sum_{im}\,\kappa_{m}\,Q_{im}^2. \nonumber
\eeq
The first term corresponds to the oxygen-mediated electron transfer between nearest-neighbor manganese sites along the $\{...x,x,y,y...\}$ chain, where $d_{i,\alpha}^\dagger$ creates a particle at site $i$ in electronic states stemming from $e_g$ orbitals $d_{x^2-y^2}\,(\alpha=a)$ and $d_{3z^2-r^2}\,(\alpha=b)$ and  $t_{aa}^{ij}=3t_0\cos^3\varphi_{ij}/4,\,t_{bb}^{ij}=t_0\cos\varphi_{ij}/4$ and $t_{ab}^{ij}=\pm\sqrt{3}t_0\cos^2\varphi_{ij}/4$~\cite{slater}, where $t_0=(pd\sigma)^2$ is taken as the bare hopping parameter. The Mn-O-Mn angle depends on the oxygen positions as $\cot(\varphi_{i\pm1}/2)=2 u_{i\pm 1/2}$, where $u_{i\pm 1/2}$ is the displacement
of the oxygen atoms around the $i$th Mn site perpendicular to the bond, see Fig.~\ref{fig1}(a).
The sign in front of the interorbital amplitude $t_{ab}^{ij}$ is determined by the direction of the vector connecting two
neighboring sites along the zigzag chain; we can distinguish then
between bridge sites, where hopping does not change direction, and corner sites, where it does. The second term describes
the Coulomb interaction, and the last terms describe the $e$-$l$
interaction, with coupling $\lambda$ and operators $n_i=d_{ai}^{\dagger}d_{ai}^{\phantom{\dagger}}+d_{bi}^{\dagger}
d_{bi}^{\phantom{\dagger}}$, $\tau^z_i=d_{ai}^{\dagger}d_{ai}^{\phantom{\dagger}}-d_{bi}^{\dagger}
d_{bi}^{\phantom{\dagger}}$ and 
$\tau^x_{i}=d_{ai}^{\dagger}d_{bi}^{\phantom{\dagger}}+d_{bi}^{\dagger}d_{ai}^{\phantom{\dagger}}$. $Q_{im}$ represents the
different phonon modes, i.e. breathing, Jahn-Teller and buckling modes, with stiffness constants $\kappa_m$. The buckling
mode on the chain corresponds to $Q_{i,\mbox{\small buckle}}= (u_{i+1/2}+u_{i-1/2}-(-1)^i\,2\,u_0)$, where $u_0$ is the bare value of the buckling distortion. We adopt standard values for the stiffness constant of the breathing and JT modes, $\kappa_1=2,\,\kappa_2=\kappa_3\equiv\kappa_{JT}=1$ and use $\kappa_{\mbox{\scriptsize buckle}} = 10$~\cite{dagotto2}. In the following we express the strength of the $e$-$l$ coupling in units of the static JT energy $E_{JT}=\lambda^2/(2\kappa_{JT}\,t_0)$. 

{\it Calculations.} We determine the ground state wavefunction variationally with respect to the phonon modes, treating the on-site interaction at the mean-field level. Since the Coulomb term appearing in Eq. (\ref{ham}) is rotationally invariant in the orbital space, the decoupling Hartree-Fock procedure is implemented in the local optimum $e_g$-electron basis, which is determined self-consistently. This allows us to retain the orbital degrees of freedom in the mean-field framework; should we linearize the interaction term in the original basis, we would fix the orbitals to be either $x^2-y^2$ or $3z^2-r^2$. The linearized interaction term is
$
4\,n_{ia}\,n_{ib} \approx 2\,(n_i\langle n_i\rangle - \tau_{z i}\langle \tau_{z i}\rangle -\tau_{x i}
\langle \tau_{z i}\rangle\,) - \langle n_i \rangle^2+\langle\tau_{z i}\rangle^2+\langle\tau_{x i}\rangle^2
$.
We first consider the $\lambda=0$ ($E_{JT}=0$) limit. When the initial buckling is small ($u_0\lesssim 0.2$), we recover the standard result for the CE-type DDEX model~\cite{dagotto.rev,brink}. At half-filling the ground state of the non-interacting system is a band insulator with bridge Mn ions displaying (partially filled) orbitally ordered states $3x^2-r^2/3y^2-r^2$, while corner Mn ions do not display any OO. As mentioned above, such a band insulating state has a topological origin which is ultimately related to the constraint on $e_g$ electron mobility introduced by the underlying magnetic configuration of $t_{2g}$ spins~\cite{brink,hotta}. Since both orbitals of the local electron-state set are partially occupied at corner sites but only one is occupied at bridge sites, the inter-orbital Coulomb interaction $U$ acts differently on them, inducing a charge transfer from
more correlated (corner) to less correlated (bridge) sites,  i.e. $U$ induces a CB charge-order pattern with Mn$^{+3.5-\delta}$/Mn$^{+3.5+\delta}$ on bridge/corner sites. As a consequence of the competition of hopping and elastic energies -- the kinetic energy is of course minimal for the undistorted structure ($\varphi =\,180\degree$) -- the optimal oxygen displacements along the chains are $u_{i\pm 1/2}=u_{eq}<u_0$.

For larger initial distortions, however, the system gains energy for physical values of $U$ by dimerizing, i.e. the optimal values of $u_{i\pm 1/2}$ become different at even and odd bonds causing an effective hopping dimerization, as shown in Figs.\ref{fig1}(a) and \ref{fig1}(b). We refer to this dimerized phase as a Zener polaron state, even if we are at this point not considering the magnetic moments of such dimers. Correspondingly, also an orbital ordering establishes at corner sites, with the most occupied orbital aligned along the direction in which the hopping amplitude is larger.  A good measure of the orbital ordering is $\Delta_{00}=\rho_{\bar{a}}-\rho_{\bar{b}}$, where $\rho_\alpha=\langle \psi^\dagger_\alpha\psi^{\phantom{\dagger}}_\alpha\rangle$ is the local average occupancy of the orbital state $\alpha$ and $\psi_\alpha=-\sin(\theta_\alpha/2)\,d_a\,+\cos(\theta_\alpha/2)\,d_b$. Here $\bar{a},\bar{b}$ label the local most and  least occupied orbital states, which we find to be $3x^2-r^2\,(3y^2-r^2)$ and $y^2-z^2\,(x^2-z^2)$ respectively. As a secondary effect, the correlation-induced charge transfer from corner to bridge sites is considerably reduced with respect to the non-dimerized state (see inset of Fig.\ref{fig1}(c)), a direct consequence of the small energy cost for an electron to partially fill a single orbital state on corner sites. This suggests that the gain in electronic energy comes from the  minimization of the on-site Coulomb energy at corner sites, directly related to the onset of the orbital ordering. The OO/CO patterns which characterize the dimerized state in the proposed model agree perfectly well with that expected in the proposed ZP picture for the half-doped manganites. We note also that the Mn-O-Mn angle depends on the nature of the bond (parallel $vs.$ perpendicular orbital states along each chain, ferromagnetic $vs.$ antiferromagnetic between different chains); in the dimerized phase we end up with three different angles, corresponding to the involved bonds, in qualitative agreement with
the PCMO measured angles~\cite{jooss}.

\begin{figure}
\includegraphics[height=4.4cm]{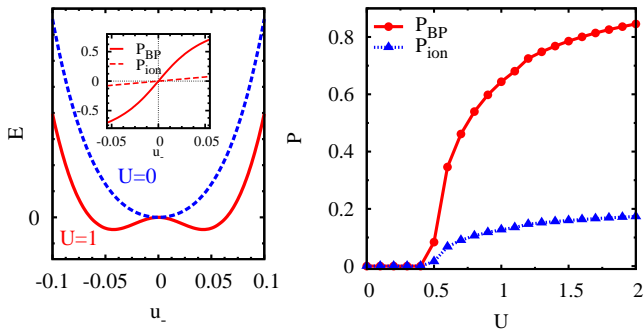}
\caption{(left) Evolution of total energy as a function of $u_{-}=(u_1-u_2)/2$ at fixed optimum $u_{+}=(u_1+u_2)/2$ at
$U=0$ and $U=1$ ($E_{JT}=0$); the inset shows the onset of the two contributions to polarization at $U=1$. (right)
Polarization evaluated at $u_0=0.3$ and $E_{JT}=0$ as a function of $U$ (the electron charge $e$ is set to unity).}
\label{fig2}
\end{figure}

{\it Multiferroicity.} Having established that buckling of the MnO$_6$ octahedra drives the system toward an instability
where the dimerized Zener polaron state can form in the presence of electron-electron in interactions, we now focus on the
possible ferroelectricity of the resulting ground state. The bridging oxygens located in the middle of the bonds are no
longer inversion-symmetry centers in the dimerized state, allowing for an in-plane ferroelectric polarization $P_{ion}$ perpendicular
to the chains direction (see Fig.\ref{fig1}(a)). In analogy with the displacement mechanism induced by the exchange coupling in
undoped rare-earth manganites, we refer to the mechanism that differentiates the bridging oxygen positions in the proposed
model as a correlation-induced orbital-striction. One can estimate $P_{ion}$ as the oxygen displacement per ``unit cell'' containing one Mn and two O, averaged over the true unit cell along the chain, consisting of four Mn sites. 
But in addition a purely electronic contribution to the polarization is generated by the interplay of the orbital order and the specific topology of the chains. In fact, $e_g$ Bloch electrons acquire a Berry phase in the OO state, and a different geometric phase coming from the change of the hopping direction~\cite{hotta,koizumi}, whose interplay gives rise to a finite polarization~\cite{mio}. In the non-dimerized AFM-CE state, the phase associated to OO appears only on bridge Mn, whereas the hopping phase due to the zigzag geometry of the chains originates at corner Mn, thus giving no polarization~\cite{mio}. On the other
hand the onset of OO on corner Mn sites allows for a finite $P$, whose magnitude is directly related to the strength of the
orbital order. We evaluated this Berry phase contribution to the polarization \cite{mio,resta2}, finding a sizable polarization parallel to the
chains direction, in excellent qualitative agreement with ab-initio calculations for both LCMO and PCMO~\cite{colizzi,gianluca}. Both the electronic and ionic contributions for the one-dimensional chain are shown in Fig.\ref{fig2}.

\begin{figure}
\includegraphics[height=3.25cm]{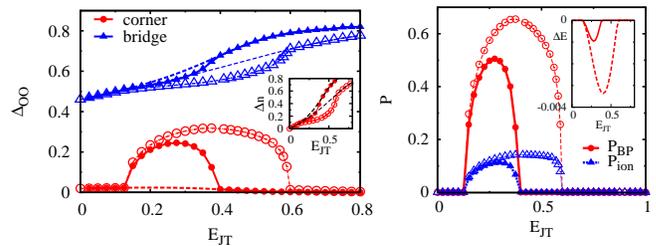}
\caption{(left) Difference in on-site orbital occupancy as a function of $E_{JT}$ in the dimerized state, on corner
(circle) and bridge (triangle) sites. Inset: charge disproportion between corner and bridge sites. Dotted lines label the
non-dimerized results. Open symbols show the same quantities for the breathing mode switched off. (right) Calculated
polarization as a function of $E_{JT}$ at $u_0=0.3,\, U=0$ (solid  and open symbols: with and without the breathing mode). Inset: energy gain of the dimerized phase with (solid line) and without (dashed line) breathing mode.} 
\label{fig3}
\end{figure}

\begin{figure}
\includegraphics[height=4.26cm]{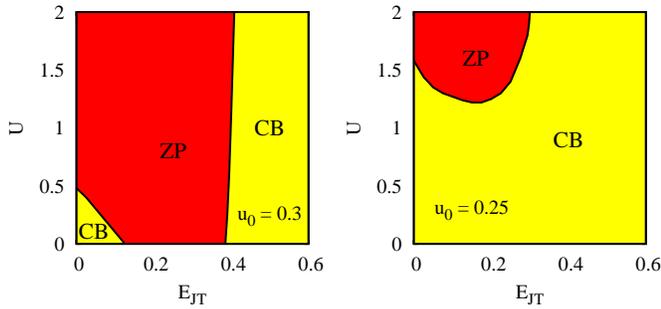}
\caption{Phase diagram in the $E_{JT}$-$U$ space at (left) $u_0=0.3$ and (right) $u_0=0.25$; ZP is the dimerized state and CB the non-dimerized one.} \label{fig4}
\end{figure}

The same analysis can be repeated in the presence of the $e$-$l$ interaction. On the basis of our previous findings, we
expect that the JT coupling, that may induce the occupation of a preferential local orbital state, would favour the onset of
the ZP dimerized phase, whereas the breathing mode would probably suppress the tendency to dimerize, favouring instead a CB
charge-ordering pattern. Indeed, at $U=0$ we find a very narrow window of values for $E_{JT}$ in which the dimerized phase
sets in, with a very small energy gain with respect to the non-dimerized state. This energy gain comes from the JT
coupling, which is thus less effective than correlation in inducing the differentiation of oxygens displacement, even if
the resulting picture is qualitatively the same (see Figs.\ref{fig3}, left, and \ref{fig3}, right). In fact,  when the breathing mode coupling is switched off the ZP stability region is significantly enlarged, a further indirect proof that the orbital ordering on corner sites plays a relevant role in the onset of the dimerized state. On the other hand, the $e$-$e$ and the JT $e$-$l$ coupling cooperate to enhance the orbital-striction mechanism, even in the presence of the breathing mode, as clearly shown in the phase diagram in the $E_{JT}$-$U$ space shown in Fig.\ref{fig4}.

{\it Conclusions.} The instability of the system toward the formation of a ZP dimerized phase depends crucially on the
competition and subtle balance between double-exchange, electron correlation and the lattice elastic energy.
Repeating the analysis above for different values for $k_{\mbox{\scriptsize buckle}}$, 
we find that the dimerization becomes weaker for smaller $k_{\mbox{\scriptsize buckle}}$, and is thus more 
stable for a stiffer lattice. Indeed,
in the first case the hopping energy gain due to the increase of $\varphi$ at both  even and odd bonds
overwhelms the energy gain that would result from these bonds becoming inequivalent;
on the other hand, the orbital-striction mechanism comes into play when the Mn-O-Mn angle cannot be increased very much at any bond.
The initial degree of buckling is essential: the more distorted the system is, the more effective the Coulomb and JT
interactions are in stabilizing ZP, see Fig.\ref{fig4}. 
In our model the magnetic degrees of freedom of the dimers and the charge degrees of freedom of the oxygens are not
included, which is justified by the observation that the most stable magnetic configuration for the half-doped case is very
close to the AFM-CE state~\cite{efremov}, even if the geometry of the exchange couplings between dimers, that is of
triangular rather than square type, can stabilize magnetic configurations different form AFM-CE~\cite{efremov}, thus
lessening the instability~\cite{note}. The oxygen charge degrees of freedom can become relevant because the enhanced
transfer energy inside each dimer can activate a charge transfer from bridging oxygens to manganeses, which is expected to
compensate for the charge disproportion between the corner and bridge Mn sites in our model. This can change the value of
the ferroelectric polarization but will not affect our conclusion on the presence of a Zener polaron instability that is induced
by the buckling and rotation of MnO$_6$ octahedra. The qualitative features of the emerging ZP state are reproduced very
well in our model Hamiltonian for manganeses only; in the present context a charge transfer from oxygen to manganese is rather the consequence of the instability.

This work has been supported by the European Community's Seventh Framework Programme FP7/2007-2013 under Grant Agreement No. 203523-BISMUTH.


\begin{thebibliography}{99}
\bibitem{note} As confirmed by DFT calculations~\cite{gianluca}, when the magnetic moments of Mn-Mn dimers
are artificially rotated, suppressing the ferroelectric polarization in the ZP state.
\end{thebibliography}
\end{document}